# "Setting things straight" by twisting and bending?

Edward T. Samulski[a]*, Alexandros G. Vanakaras[b], and Demetri J. Photinos[b]

[a] *Department of Chemistry, University of North Carolina, USA;* [b] *Department of Materials Science, University of Patras, Patras, Greece*

email: et@unc.edu

**Abstract**. We explain the origins of the controversy about the classification of the $N_X$ phase observed in cyanobiphenyl dimers and why it is a polar twisted phase, an entirely new kind of nematic.

## I. Introduction

Last year we decided to confront the pervasive and errant $N_{TB}$ classification of the second, lower temperature nematic phase of cyanobiphenyl dimers (CBnCB dimers with an odd number n -CH$_2$- units in the aliphatic linker) denoted $N_X$. Our Article, "The Twist Bend Nematic: A Case of Mistaken Identity" [1](for brevity SVP), was both a critique of the putative $N_{TB}$ discovery and a reiteration of an earlier proposal for the actual supramolecular structure in the $N_X$—the polar twisted nematic phase $N_{PT}$. Our submission was met with an unusual editorial decision: *Liquid Crystals* stipulated that our Article would only be published if accompanied by a rebuttal of our proposal ($N_X \neq N_{TB}$), intended to "counterbalance" the content of our Article. Herein we reply to that rebuttal, "Setting things straight in 'The twist-bend nematic: A case of mistaken identity.'" , an "Invited Comment" by Dozov and Luckhurst, [2] (DL, for short).

As stated at the outset, our goal was to engender a better understanding of the structure of the unusual, second nematic phase of odd-linked CBnCB dimers. We were motivated to do that in part because the $N_X$ had been misidentified several times since its discovery in 1991. Intially it was believed to be a biaxial nematic [3,4] then a smectic [5], then a nematic again [6], and eventually in 2011 [7] a *twist bend nematic* ($N_{TB}$) , a phase proposed by Robert Meyer in 1973 [8]. Finally in 2016, two of us (Vanakaras and Photinos) proposed that the structure in the $N_X$ was an entirely new kind of nematic, the $N_{PT}$ [9].

Dozov and Luckhurst do not address the actual contents of our Article [1] but present an incorrect interpretation of it and the discoveries that led up to the $N_{PT}$ model. We respond to their Comment by explaining here the fundamentals of elasticity that underlie our position, and by addressing their inaccuracies point by point in an Appendix that culminates with a history of the $N_X$ phase and its modeling. We differentiate between descriptions of the nematic phase generated on the basis of the Frank-Oseen elasticity theory and those based on molecular models. This fundamental distinction is at the core of the controversy about the classification of the $N_X$ phase. To facilitate reading, citations with the prefixes SVP or DL correspond to references with the numbering used in the original works [1] and [2] respectively. Where verbatim quotations are used, we have kept unchanged the original numbering and style of references.

**II. Fundamentals of elasticity in nematics**

The Frank-Oseen continuum theory (F-OCT) describes the elastic behavior of nematics, the basis of Robert Meyer's twist-bend nematic conjecture. Many texts have reviewed this theory and concur about its centrality to liquid crystal behavior. In particular, De Gennes' and Prost's text[10] section "3.1 PRINCIPLES OF CONTINUUM THEORY", shows that the Frank-Oseen elasticity theory is restricted to i) deformations over regions of spatial dimension $l$ much larger than the molecular dimensions $a$ ($a/l \ll 1$), ii) to "soft" curvatures of the nematic director ($a\nabla \mathbf{n} \ll 1.$), and iii) to the preservation of the the $D_{\infty h}$ local symmetry of the molecular ordering. These restrictions allows for the definition of **n** and its curvatures, which in turn specifies the possible deformations (bend, splay and twist) of the director field. On the basis of these considerations, some questions, regarding deformations described in the framework of the Frank-Oseen continuum theory (F-OCT), can be readily addressed.

*Is it allowable to treat deformations taking place on a molecular length scale?*

No, because this would be inconsistent with $a/l \ll 1$ and therefore also at variance with the requisite neglect of "terms of higher order in $(a/l)$" in eq (3.2) of [10] (see section 5 in the Appendix). Dozov and Luchkurst in their Comment argue that:

> "Indeed, the scalar order parameter, $S(\mathbf{r})$, and the director, $\mathbf{n}(\mathbf{r})$, are defined by averaging the orientation of all the molecules in some volume, $V_{av}$, around the position **r**. If the director varies rapidly in the space, as is the case in the $N_{TB}$ phase, it might be difficult to choose a sufficiently large volume…. However, due to the 1-dimensional distortion of the $N_{TB}$ phase, $V_{av}$ may be as large as required if we choose it as a thin ($\ll p$) slice perpendicular to the helix axis…."

The issue is not the mere definition of the director in some volume *v*, but the deformation of **n(r)**, described by the curvatures of the director field, i.e., it has to do with lengths. In order to describe the curvatures of **n(r)**, the director has to be definable over a small volume *v* around **r**, and of course such a description is meaningful only if the length scale of the curvature of **n**(r) is much larger than the dimensions of *v*. Obviously, if the curvatures of **n(r)** vanish within a thin ($\ll p$) slice, then *v* can be taken to be the volume of the entire slice. Irrespective, however, of how thin such a slice is, consistency with the F-OCT requires the length scale (*l*) of the director field curvature to be much larger than the molecular dimensions: $a \ll l \Rightarrow p \gg a$; i.e. the pitch is necessarily much larger than molecular dimensions.

Dozov and Luckhurst continue:

> "Samulski *et al*. state in a recurrent manner that the pitch predicted by R.B. Meyer is $\gg 1$ μm. This statement is completely wrong ‐ Meyer [10] did not even mention the scale for the pitch of the $N_{TB}$ phase that he had proposed. … The expected pitch can vary between large limits ‐ from $p = \infty$ (for lath-like molecules) to, maybe, $p \ll 100$ nm (for highly bent molecules similar to those that form the experimentally observed $N_{TB}$ phase)."

First, figure 1 of our Article shows our view of the structure and local symmetry, not only of the $N_{TB}$ proposed by R.B.Meyer, but of <u>*any model phase that is formulated in the framework of the F-OCT and yields a stable state with spontaneous twist and bend deformations of the director field* **n(r)**</u>. As discussed above, a condition for the applicability of the F-OCT is that the length scale of the director curvatures is much larger than the molecular dimensions, $a/l \ll 1$. Given that in this case the relevant curvature length is identified with the helical pitch *p*, and that the typical length scale of director elastic deformations is on the μm scale, we indicated in figure 1 that the pitch is within that range.

Second, Meyer said: "As we shall see later, the small torque produced by a magnetic field acting on the anisotropy of the magnetic susceptibility is sufficient to produce curvature strains with a radius of curvature of only a few microns." (page 285 of [8]; DL[10]) By the time he reached

page 319 where he introduced his twist-bend conjecture, Meyer had described several examples of elastic deformations and their respective length scales.

Third, Dozov and Luckhurst admit some concern about the incompatibility between the measured $N_X$ pitch and the length scale conditions for the consistent applicability of the F-OCT, albeit indirectly, in this statement:

> "At the time of these predictions even a 50-100 nm pitch was considered as unrealistically small and this resulted in an almost unanimous disbelief in the existence of the predicted phase. Nobody, not even the authors of the predictions, had forecast that the actual pitch of the $N_{TB}$ phase is even smaller, typically <10 nm."

### *Is it possible to treat polar or biaxial nematic phases within the F-OCT ?*

The answer is again clearly negative. A lower than uniaxial local symmetry implies that the order parameter in eq (3.2) (section 5 in the Appendix) is not the sole order parameter of the medium, even if **n** remains a local symmetry axis; in any case the symmetry conditions leading to the free energy expression in eq (3.15) (section 5 in the Appendix) are no longer valid. Therefore, irrespective of the assumed <u>molecular</u> symmetry, consistency requirements impose local uniaxial symmetry for any spontaneously deformed nematic phase formulated in the framework of the F-OCT. This is precisely depicted in figure 1 of SVP.

Naturally, a bend deformation introduces the bend vector **b,** perpendicular to **n;** similarly, a twist deformation introduces a pseudoscalar *t*. The condition $a\nabla \mathbf{n} \ll 1.$ in section 6 of the Appendix implies both $a|\mathbf{b}|\ll 1$ and $a|t|\ll 1$. Therefore, the polarity introduced by the bend structure has a necessarily negligible effect on the local symmetry of the molecular ordering: the polarity induced in the local molecular ordering by the bend deformation **b** is on the order of $a|\mathbf{b}|$ or smaller; we referred to that as "phantom polarity." Analogously, a twist deformation would induce a chiral effect on the molecular conformations, whose magnitude is on the order of $a|t|$ or smaller, and therefore negligible.

The deformation-induced asymmetry and local symmetry of the molecular organization are often conflated in Dozov's and Luckhurst's Comment and in other works by these authors, as in these examples:

> " The low symmetry has been discussed also when the polar order results from the bend distortion of the director: … The flexoelectric coupling of the local vector order with molecular dipoles results in macroscopic electric polarization of the phase [12]."
>
> "…this symmetry breaking can be expected also in the absence of smectic order, inside the nematic phase. … If further experimental studies confirm this prediction, one can expect nematic phases with low symmetry and large spontaneous electric polarization [12]."
>
> "The local biaxial order of the NTB phase has attracted less attention in the early models, but is obvious in all of them."

Referring to the $N_{TB}$ model of Dozov, (DL[12]), which has the F-OTC as its starting point, the authors speak of the coupling between the bend-induced polar order with some "local vector order" which further gives rise to macroscopic electric polarization. Then they refer to the "obvious" local biaxial order even though the F-OCT formulation precludes that symmetry.

In discussing a more recent member of their large $N_{TB}$ family of models, they finally understand the local symmetry of the $N_X$—albeit under the pseudonym $N_{TB}$—some years after Vanakaras and Photinos (2016) published a detailed description of their model [9] (DL[15]):

> "Finally, the relations between the molecular, local and macroscopic symmetries were discussed in detail in the coarse-grained model of the $N_{TB}$ phase developed by C. Meyer and I. Dozov."
>
> "Due to the low symmetry of the bent-shape molecules and to the strong heliconical distortion of the director, the $N_{TB}$ phase has very low local symmetry, with additional biaxial and polar long-range orientational order [35]."
>
> "On a length scale much shorter than the $N_{TB}$ heliconical pitch, $\Lambda \ll p$, the symmetry of the phase is even lower, $C_2$. … This symmetry is much lower than the symmetry of the nematic ($D_{\infty h}$), SmA ($D_{\infty h}$) or SmA* ($D_\infty$) phases. For this reason, we expect that the local properties of the $N_{TB}$ phase will be substantially different from the properties of those more symmetric phases. In particular, at this microscopic scale, the distortion energy of the twist-bend nematic should be different from the usual Frank elastic energy. We note that due to the low local symmetry, the usual nematic uniaxial tensor order parameter, **Q**, is not sufficient for the description of the $N_{TB}$ phase: the order tensor becomes slightly biaxial, as in the N* phase, and there is also an additional polar orientational order [37]."

What they are describing here is exactly the symmetry in the $N_{PT}$ illustrated in figure 2 of our Article. They also correctly indicate that under such symmetry conditions the F-OTC approach is not applicable. This means, however, admitting that all of the $N_{TB}$ models obtained within the F-OTC context (including the initial Dozov model DL [12]) i) do not (and could not) describe the correct $N_X$ local symmetry, and ii) if endowed *a posteriori* with polar molecular ordering, they would be directly violating the validity of the conditions upon which their derivation is based.

Nevertheless, they insist that *we* do not understand.

> "Let us now consider why Samulski *et al*. seem not to understand that their newly invented "$N_X$" phase is indeed the $N_{TB}$ phase and further identify it as being the $N_{PT}$ phase. The answer to the first question can be found in Fig. 1 of Ref. [18]. This figure shows the $N_{TB}$ phase as it is imagined by Samulski *et al*. and demonstrates their deep and surprising misunderstanding of the twist-bend nematic phase.

It is true that we do not understand how they can occupy what appear to be contradictory positions on the subject. In summary, any $N_{TB}$ model variant obtained within the framework of F-OCT is "by construction" restricted to: i) uniaxial and apolar symmetry of the local molecular ordering and, ii) a pitch for the heliconical modulation length that is much larger than molecular dimensions. In their disregard of the consistency requirements for the F-OCT, Dozov and Luckhurst fail to acknowledge that of all the models they subsume into their $N_{TB}$ family cluster, the only ones that directly and naturally lead to a molecular length scale pitch and to polarity of the local molecular ordering, i.e., those in their Refs DL[11] and DL[15], are precisely the models that are developed entirely outside of the framework of the F-OCT.

After the above considerations, it becomes clear that in a physically discriminating classification, the $N_{TB}$ model is understood to originate conceptually from the simultaneous presence of two of the three elastic deformations that the nematic director field **n(r)** can assume. This automatically imposes the length scale of the modulation and the local symmetry of the molecular ordering. Indeed, both the original R.B. Meyer formulation and the subsequent Dozov formulation are based explicitly on the Frank-Oseen continuum theory. In the model of Meyer, the elastic free energy is extended to encompass flexoelectric and flexopolarisation couplings; such extension was shown possible to stabilize states with spontaneous twist and bend elastic deformations of the director field **n(r)**. The same elastic deformations were obtained by Dozov in his formulation, wherein the case of a negative bend elastic constant ($K_3$ in eq(3.15) of section 6 in the Appendix) is considered. In this case, the Frank-Oseen free energy is extended to include *ad hoc* quadratic second derivative terms of the director field $\mathbf{n}(\mathbf{r})$ which confer stability to the spontaneously deformed states.

Clearly, modulations on the molecular scale are not physically related to elastic deformations of the director field $\mathbf{n}(\mathbf{r})$ and therefore a consistent description cannot be obtained in the context of nematic elasticity. On the other hand, molecular theories and molecular simulations are not subject to the limitations (length-scale and local symmetry) of elastic continuum formulations and are therefore the proper tools to use for consistently accounting for such modulations. The $N_{PT}$ is one such example wherein the molecular structure mimics the essential features of the odd-spacer dimers albeit in a very simplified way (hence the designation "toy-model"). Within that description the polar ordering of the molecules emerges naturally and is seen to drive the molecular length scale modulation; hence the attributes of "polar" and "twisted." Here, "twisted" is not to be confused with the "twist elastic deformation" of the director field $\mathbf{n}(\mathbf{r})$, since it is neither an elastic deformation nor does it refer to the the nematic director $\mathbf{n}$. An alternative, more rigorous term for this molecular-scale modulation would be "roto-translation," thereby eliminating any possible confusion with the elastic twist deformation. These clear physical distinctions between the $N_{TB}$ model and the $N_{PT}$ model are summarized in Table 1.

**Table 1.** Physical characteristics that differentiate the $N_{TB}$ model from the $N_{TP}$ model.

| **General Aspects** | **Attributes** | **$N_{TB}$** | **$N_{PT}$** |
|---|---|---|---|
| Local Symmetry | Symmetry Group | $D_{\infty h}$ | $C_2$ |
| | Symmetry Axis | Nematic Director $\mathbf{n}(\leftrightarrow -\mathbf{n})$ | Polar Director $\mathbf{m}(\not\leftrightarrow -\mathbf{m})$ |
| Spatial Modulation | Type | Twist & Bend of $\mathbf{n}$ | Roto-translation of $\mathbf{m}$ |
| | Length Scale | Macroscopic $p \sim \mu m$ | Molecular $p \sim nm$ |
| | Physical Origin | Spontaneous Elastic Deformations of $\mathbf{n}$ | Polar Molecular Organization along $\mathbf{m}$ |

**III. Conclusions.**

After reading carefully the Comment by Dozov and Luchurst, we could not find in it anything that would prompt us to reconsider or even rethink what we have written in our Article. We feel, however, that our effort to read and try to understand their Comment did not go unrewarded, although in a somewhat unexpected way:

First, in their rebuttal Comment the authors in effect confirm and corroborate two essential points made in our Article: i) one regarding the local symmetry and modulation of the lower temperature $N_X$ phase in dimers, ii) the other regarding the incompatibility of these features with the original $N_{TB}$ modeling despite their subsequent revisions of that modeling in order to salvage the identification of $N_X$ with $N_{TB}$.

Second, we now realize that all along Dozov and Luckhurst have been the prime contributors to the above revisions of $N_{TB}$, and in fact their rebuttal is a culmination of that. Although they insist on retaining the name $N_{TB}$, it is clear that the symmetries and modulations of the $N_X$ phase have nothing to do with any elastic twist and bend deformations of the nematic director. As a result Meyer's model of the $N_{TB}$ gets subsumed into an ill-defined cluster of twist bend nematics— and Dosov's also.

The initial aims of our Article remain the same: i) to highlight the basic errors propagated by Dozov, Luckhurst et al. in their $N_{TB}$ modeling of the $N_X$ phase of CBnCB dimers, and ii) to extricate Robert Meyer's original conjecture from an ever-expanding and dysfunctional family of twist bend nematic models, so that a new generation of researchers will not be discouraged from searching for the yet-to-be-discovered $N_{TB}$ phase.

In the end, we wish to thank Dozov and Luckhurst and all of the scientists who have contributed energy and time to this subject for the past three decades. While we cannot agree with all of their conclusions, their continuing interest and efforts have helped us to hone our perspective and to reach a new understanding of the underlying physics of the $N_X$ phase.

> "A round man cannot be expected to fit in a square hole right away.
> He must have time to modify his shape."     – Mark Twain

**References**

[1]  Samulski ET, Vanakaras AG, Photinos DJ. The Twist Bend Nematic: A Case of Mistaken Identity. Liq Cryst. 2020; https://doi.org/10.1080/02678292.2020.1795943

[2]  Dozov I, Luckhurst GR. Setting things straight in 'The twist-bend nematic: A case of mistaken identity.' Liq Cryst. 2020; https://doi.org/10.1080/02678292.2020.1795944

[3] Toriumi H, Kimura T, Watanabe H. Alkyl Chain Parity Effect in the Phase Transition Behavior of a,ω-Bis(4,4'-cyanobiphenyl)alkane Dimer Liquid Crystals. Sumbitted for publication in Liq Cryst. 1991;

[4] Emerson APJ, Luckhurst GR. On the relative propensities of ether and methylene linkages for liquid crystal formation in calamitics. Liq Cryst. 1991;10:861–868.

[5] Barnes PJ, Douglass AG, Heeks SK, et al. An enhanced odd-even effect of liquid crystal dimers Orientational order in the α,ω-bis(4′-cyanobiphenyl-4-yl)alkanes. Liq Cryst. 1993;13:603–613.

[6] Panov VP, Nagaraj M, Vij JK, et al. Spontaneous Periodic Deformations in Nonchiral Planar-Aligned Bimesogens with a Nematic-Nematic Transition and a Negative Elastic Constant. Phys Rev Lett. 2010;105:167801.

[7] Cestari M, Diez-Berart S, Dunmur DA, et al. Phase behavior and properties of the liquid-crystal dimer 1″,7″-bis(4-cyanobiphenyl-4′-yl) heptane: A twist-bend nematic liquid crystal. Phys Rev E. 2011;84:031704.

[8] Meyer RB. Structural Problems in Liquid Crystal Physics. Mol Fluids. Gordon and Breach, New York; 1976.

[9] Vanakaras AG, Photinos DJ. A molecular theory of nematic–nematic phase transitions in mesogenic dimers. Soft Matter. 2016;12:2208–2220.

[10] Gennes PG de, Prost J. The Physics of Liquid Crystals. Oxford: Clarendon Press; 1995.

[11] Memmer R. Liquid crystal phases of achiral banana-shaped molecules: a computer simulation study. Liq Cryst. 2002;29:483–496.

[12] Hoffmann A, Vanakaras AG, Kohlmeier A, et al. On the structure of the Nx phase of symmetric dimers: inferences from NMR. Soft Matter. 2015;11:850–855.

[13] Kumar A, Vanakaras AG, Photinos DJ. Polar Molecular Ordering in the NX Phase of Bimesogens and Enantiotopic Discrimination in the NMR Spectra of Rigid Prochiral Solutes. J Phys Chem B. 2017;121:10689–10703.

[14] Vanakaras AG, Photinos DJ. Molecular dynamics simulations of nematic phases formed by cyano-biphenyl dimers. Liq Cryst. 2018;45:2184–2196.

[15] Palermo MF, Pizzirusso A, Muccioli L, et al. An atomistic description of the nematic and smectic phases of 4-n-octyl-4′ cyanobiphenyl (8CB). J Chem Phys. 2013;138:204901.

[16] Chen D, Porada JH, Hooper JB, et al. Chiral heliconical ground state of nanoscale pitch in a nematic liquid crystal of achiral molecular dimers. Proc Natl Acad Sci. 2013;110:15931–15936.

[17] Hori K, Iimuro M, Nakao A, et al. Conformational diversity of symmetric dimer mesogens, α,ω-bis(4,4′-cyanobiphenyl)octane, -nonane, α,ω-bis(4-cyanobiphenyl-4′-yloxycarbonyl)propane, and -hexane in crystal structures. J Mol Struct. 2004;699:23–29.

[18] Heist LM, Samulski ET, Welch C, et al. Probing molecular ordering in the nematic phases of para-linked bimesogen dimers through NMR studies of flexible prochiral solutes. Liq Cryst [Internet]. 2020 [cited 2020 Feb 15]; Available from: https://doi.org/10.1080/02678292.2019.1711214.

**Appendix - Inaccuracies in the Invited Comment by Dozov and Luckhurst**

*1. Inaccurate statements about the contents of our Article (SVP).*

The verbatim comments in the Dozov-Luckhurst Comment include:

> "We show that, contrary to the claim of Samulski *et al*., there is not a unique $N_{TB}$ model, but several different and independently developed models."

> "They state that the only $N_{TB}$ model is that of R. B. Meyer…"

> "contrary to the claims of Samulski *et al*., there is not a unique and universal $N_{TB}$ model. Instead, in the last fifty years a large number of distinct and often independent models have been proposed that predicted or described *a posteriori* the $N_{TB}$ phase."

We do explicitly and repeatedly refer to their evolving modeling, e.g., "Subsequent variants,…", "More recent attempts, with the same aim to reinstate $N_{TB} = N_X$,…", "… such revisions of the $N_{TB}$ model…", "…Meyer's elegant and physically clear <u>original</u> proposal.. .", "As a result of such continuing revisions of the $N_{TB}$ model… the <u>original</u> notion of the $N_{TB}$ is becoming obscure…." etc. Dozov and Luckhurst also incorrectly say,

> "Yet another claim, that the "$N_X$" phase cannot be described by the $N_{TB}$ models, because *it is not nematic* (?), has apolar order, and is locally biaxial, is also false:…"

> "Samulski *et al.* state that the "$N_X$" phase…is not a nematic phase at all because of its lower symmetry, lack of "apolar director" and short pitch."

We nowhere claimed that the $N_X$ phase is not a nematic. The nematic nature of the phase is repeatedly and clearly referred to, e.g., "…a more thorough study showed that the low temperature phase was nematic thereby increasing interest in this second nematic phase. That low temperature nematic was designated $N_X$", "what is the nature of the lower temperature nematic phase, $N_X$", "All of the key attributes of the $N_X$ phase are readily accounted for by a new nematic phase model, the polar twisted nematic ($N_{PT}$)", "… has prevented a clear picture of this new nematic state from emerging.", "…the organization that applies in the $N_X$, the lower temperature nematic phase of the odd CBnCB dimers (Figure 2)," etc. They also inaccurately attribute a statement to us,

> "But the cherry on the cake is their statement that the small value of the pitch for the "$N_X$"
> phase <u>even calls into question its very classification as a nematic</u>".

Immediately after quoting a passage from de Gennes, our Article refers to issues raised on the classification of nematics by Goodby, Mandle and coworkers: "…<u>and even calls into question its very classification as a nematic</u> [24, 28] …". This quotation is not a claim by us that the Nx is not a nematic. We give its origins by citing the corresponding sources but those references are omitted from their Comment. For example, in SVP[28] it is noted that, "Previously we suggested the phase be described as 'twist-bend' rather than 'twist-bend nematic' (i.e. TB not $N_{TB}$)[22] as the local helical structure of this phase is at odds with the definition of a nematic phase as defined by IUPAC, sharing only the lack of positional order of a true nematic phase."

> "Samulski *et al.* state that it cannot be the predicted $N_{TB}$ phase, because the $N_{TB}$ phase is
> *nematic*. Instead, they claim that the *nematic* "$N_X$" phase is in fact the $N_{PT}$ phase, which is
> (as they state) *not nematic* (despite its name, "polar-twisted *nematic*")."

At best these statements are "constructions;" they are nowhere made or implied by us. The misidentification of $N_X$ as $N_{TB}$ has nothing to do with the $N_{TB}$ just being a nematic. Another example,

> "Memmer's Monte Carlo, … Clearly during the simulation the box dimensions are not
> constant as implied by Samulski et al.[18]."

We never implied that Memmer's box dimensions are constant. Rather we said, "But Memmer's pitch scale was determined *a priori* by merely setting it equal to the simulation box length;" and this conveys precisely what is specified in Memmers paper [11]: "… a length of the pitch has been formed inside the simulation box, i.e. pitch $P^* = L_z^*$"

## 2. *Inaccuracies about the contents of papers concerning the $N_{PT}$ model.*

> "However, it was subsequently demonstrated that the analysis of the NMR spectra in Ref.
> [14] was incorrect because it had ignored the fact that translational diffusion of the molecules
> along the helix axis could remove the effect of the phase biaxiality on the NMR spectrum
> [64]."

This fact was not ignored. It was addressed directly and discussed in detail in ref [12] (DL[14]):

"A second possibility would be to assume that the translational diffusion of entire dimer molecules … is rapid on the NMR time scale (in this case $10^{-4}$ s); then the time-averaged molecular motions, …, would produce an effectively uniaxial spectrum about the helix axis…. To our knowledge there is no quantitative evidence to directly prove or disprove such assumptions." Moreover, we clearly pointed out the implications in case of rapid diffusion: "However, if such assumption is proven valid it would simply mean that the NMR techniques used here and in (Ref. 8, 9 and 16), are "blind" to the possible presence of a heliconical twist-bend structure of the director in the $N_X$ phase."

> "The next variant of the same model appeared in Ref. [15], under the name "toy-model" for the "$N_X$" phase. The structure proposed here is exactly the same as in the previous paper."

The paper in DL[14] was on NMR *experiments* in the $N_X$ and *not* about modeling. Therefore DL[15] (Ref. [9] here) could not be described as a variant of any model, given that the Vanakaras-Photions model did not exist at that point. An attempt to interpret the experimental findings was made in DL[14]. The inadequacy of the limited set of findings prevented the formulation of a model and that was clearly mentioned in DL[14]: "As the NMR technique used is not sensitive to the distribution of molecular positions and the data presented here are from just the deuterated termini of the dimer spacer, no firm inferences can be drawn on the detailed molecular organization within the chiral domains of the $N_X$ phase. An extensive consideration of the possible structures is undertaken in a forthcoming communication[27] based on a wide set of available observations." That "forthcoming communication" is precisely Ref DL[15] wherein it is stated that, "A detailed molecular picture of the structure within these domains, which of course could not be obtained solely from the NMR experiments, is consistently provided by the molecular theory presented in this work."

> "The structure proposed here is exactly the same as in the previous paper, with the main axis of the bent-shaped dimer molecule, i. e. the axis parallel to the all-trans configuration of the alkyl-chain spacer, oriented parallel to the pitch axis … This mechanism does not induce any change in the nematic structure and symmetry of the "toy-model" phase: the nematic director remains uniform, as in the $N_U$ phase, …"

> "Although the model is cited as being the same as that from Ref. [15], in both these papers appears, in a somewhat hidden manner, a drastically new feature which is absent in the "toy-

> model" - the main axis, **z**, of the dimer molecule (the axis connecting the centres of the two mesogenic units), which in Ref. [15] is on average parallel to the helix axis, **h**, in Refs. [16, 17] is on average tilted at some angle to **h**: The helical twisting of the polar director forces the principal axis of the ordering tensor for the mesogenic units to also twist, albeit forming a fixed angle with the helix axis (the "tilt" angle)."

First, there is no nematic director in the "toy-model" proposed for the structure of the Nx and this is thoroughly discussed and explained in DL[15]. Hence the insistence that this non-existent director remains uniform is nonsense.

Second, the **z** axis is not on average parallel to the **h** axis in the "toy model" of the Nx phase. This is thoroughly discussed in section 4.4 of DL[15] and the deviation angle of the **z**- ordering from the helix axis **h** is explicitly plotted in figure 5 of DL[15] (as $\tilde{\theta}^{(d)}$, together with the respective angle, $\tilde{\theta}^{(L)}$, for the mesogenic units). Moreover, the text just below the caption of the same figure 5 reads, "Following the twisting of the director **m**, the principal axes $\tilde{\mathbf{n}}^{(L)}$ and $\tilde{\mathbf{l}}^{(L)}$ also twist on moving along the helical axis $\mathbf{n}_h \parallel Z$, maintaining constant ''cone angles'' $\tilde{\theta}^{(L)}$ and $\tilde{\theta}^{(L)} + \pi/2$, respectively, with the helical axis. The tilted twisting of $\tilde{\mathbf{n}}^{(L)}$ bears a resemblance to the heliconical arrangement of the nematic director **n** in the twist-bend model, $N_{TB}$, of the nematic phase.[23,24] However, the resemblance is only superficial because, unlike the nematic director **n**, the principal axis $\tilde{\mathbf{n}}^{(L)}$ is not a symmetry axis, global or local." Their statements "a somewhat hidden manner" and "a drastically new feature which is absent in the "toy-model" reflect a misreading of the cited works. Consequently, their statement, "We note that in the absence of tilt, as assumed in the "toy-model", the pair correlation function of the **z** axis will be flat …" is also false.

> The next variant of supposedly the same model is reported in Ref. [16] (omitting the earlier "toy-model" denomination) and Ref. [17] (renamed to $N_{PT}$ model).

Neither reference reports any variants of the same model. DL[16] uses the model in DL[15] to describe enantiotipic discrimination measurements in the Nx phase, as stated in the respective introduction of DL[16]: "Here, on the basis of the symmetry of the environment sensed by a rigid solute molecule in the $N_X$ phase, derived from the molecular model in ref 19, we obtain the general form of the potential of mean torque that governs the orientational distribution of such molecule in the $N_X$ phase, which in turn determines the structure of the respective NMR spectra." [13]. On the other hand, DL[17] is a paper on molecular simulations, as is apparent from the title and from the abstract: "Molecular dynamics simulations of selected members of the cyano-biphenyl series of dimers (CBnCB) have been set up using atomistic detail interactions among intermolecular pairs of united atoms and allowing fully for the flexibility of the spacer chain" [14]. The input is a standard force field; no modelling is involved at any stage, only the testing of the model formulated in DL[15]. As mentioned explicitly in the abstract of DL[17]: "Key findings of the simulation are shown to be correctly predicted by the theoretical model of the polar-twisted nematic ($N_{PT}$) phase [A.G. Vanakaras, D.J. Photinos, Soft Matter. 12 (2016) 2208–2220]."In DL[17], the phase denoted as $N_{PT}$ is directly and unambiguously identified with the polar-twisted nematic phase developed in DL[15] on the basis of *a toy model for the dimer molecules*; there are no "variants." The denomination $N_{PT}$ was subsequently used for practical reasons, in place of the polar twisted nematic phase model introduced in [9] (DL[15]).

Eventually, Dozov and Luckhurst pose the question,

> "Is the structure proposed for the $N_{PT}$ model different from that of the $N_{TB}$ model?"

and answer it as follows,

> "Therefore, in contrast to the earlier 'toy model', the later $N_{PT}$ model describes, in fact, the twist-bend nematic phase and is, therefore, a $N_{TB}$ model (despite the different terminology used by the Authors of the $N_{PT}$ model)."

However, a detailed discussion of the "Apparent similarities and important differences from the nematic twist-bend model" is given in section 4.4 of DL[15]. There it is clearly shown that the twisting of the polar director gives rise to heliconical modulations of various principal axes; it is also explained in detail that such modulations do not imply equivalence with the $N_{TB}$ model.

Notably, in SVP we had said that through a sequence of evolving versions of the $N_{TB}$ their authors ended up with the $N_{PT}$ but insist on calling it $N_{TB}$. This is now reversed in DL:

> "Paradoxically, Samulski *et al.* now agree completely with this structure of the phase although curiously using a different name to refer to it (see below our analysis of the NPT model)"

Of course "this structure" in the quote from DL was only appreciated by those authors long after the "toy model" $N_{PT}$ of Vanakaras and Photinos was published DL[15].

Lastly,

> "The Originators of the $N_{PT}$ model and the computer simulations supporting it, take directly the coarse-grained averages and so define only the coarse-grained order parameters, avoiding the intermediary average at fixed Z."

The three graphs in Fig 6 of DL[17] show the explicit Z-dependence of various quantities obtained in the simulation. In fact, the main body of the results and discussion presented in ref DL[17] are devoted to detailed calculations of the local molecular ordering in the $N_X$ phase and the corresponding local "director-frame," within thin slabs perpendicular to the helical axis, and of the orientational correlations between the slabs as a function of their separation Z (see section 4, eqs (5),(8) and Fig.6(a-c) in DL[17]). The main finding of this analysis is the existence of polar molecular ordering within the slabs, defining uniquely the local polar director. This polar director is a local $C_2$ axis twisting tightly at right angles to the helical axis with a constant pitch of about of 8 nm.

## 3. Evolution of the DL "$N_{TB}$ model" to conform to $N_X$ properties.

In our Article we said, "As a result of such continuing revisions of the $N_{TB}$ model to account for new experimental data in the $N_X$ phase, the original notion of the $N_{TB}$ is becoming obscure…" This is spectacularly confirmed by the multiple statements in the Dozov Luckhurst Comment:

> "In reality, there is no unique $N_{TB}$ model but a cluster of different models that have been proposed before the experimental identification of the phase indeed new $N_{TB}$ theories continue to be developed. These models are based on different physical assumptions and lead to more or less different predictions for the physical properties of the phase and their relation with the molecular structure."

> "…in the last fifty years a large number of distinct and often independent models have been proposed that predicted or described *a posteriori* the $N_{TB}$ phase."

> "The low symmetry of the twist-bend nematic has been discussed explicitly in practically all

> the models of the N$_{TB}$ phase since their proposal. This low symmetry is obvious when a polar order parameter (primary [11] or secondary [10]) is introduced explicitly."

> "The predictive power of the $_{NTB}$ models was further confirmed by Greco *et al.* [27], …This advanced N$_{TB}$ model predicted …, and confirmed the local polar order, the tilted director, and the spontaneous twist-bend distortion of the N$_{TB}$ phase,…Using realistic molecular parameters, this advanced N$_{TB}$ model predicted, deep in the N$_{TB}$ phase, very small values of the pitch, only 2-3 times the full dimer length. …in other well-developed N$_{TB}$ models [12, 25, 26, 31]".

> "As in any of the other N$_{TB}$ models, the N$_{PT}$ model has its specific features differentiating it from other N$_{TB}$ models."

Dozov and Luckhurst evolve and broaden the notion of the twist-bend nematic claiming for the Lorman-Mettout C-phase (Ref DL[11]) that,

> "*a posteriori* it is clear that the structure of the C-phase is the same as that of the N$_{TB}$ phase",

without explaining what it is that "bends" in this C-phase, despite saying that,

> "Lorman and Mettout considered these phases as new mesophases [11], different from the smectics and nematics."

In their broadening of the definition of the "cluster" of N$_{TB}$ models/phases, they incorporate the proposed N$_{PT}$ declaring it to be

> "…a carbon copy of the N$_{TB}$ phase."

Finally, to ensure that no nematic is left out of their "cluster" of different models they say,

> "As the Meyer model is the only N$_{TB}$ model which does not postulate highly bent molecules, the expected pitch can vary between large limits – from $p = \infty$ (for lathe-like molecules) to, maybe, $p \ll 100$ nm."

thereby including the N$_U$ phase in their cluster of N$_{TB}$ models.

The collection of everything into this "cluster" of N$_{TB}$ models leads them to an unsupportable conclusion:

> "Consequently, our present analysis confirmed unambiguously that the "$N_X$" phase is indeed the $N_{TB}$ phase,"

preempting an obvious question; "Which of all these $N_{TB}$ models unambiguously corresponds to the $N_X$ phase?" They can only conclude,

> "What remains still an open and actively discussed question is which one of the $N_{TB}$ models better matches the experimentally observed $N_{TB}$ phase formed by bent bimesogens."

Searching for the criteria employed by Dozov and Luckhurst to include a nematic model in their cluster of $N_{TB}$s, we find this statement:

> "The common features of these models, which define them as belonging to the $N_{TB}$ family, is that they describe a phase with the same heliconical $N_{TB}$ structure, namely a nematic, with spontaneous bend distortion and doubly degenerate (when formed by achiral molecules) conical helix having a short pitch. Despite this similitude, the main $N_{TB}$ models imply different physical mechanisms and predict different physical properties of the twist-bend nematic phase."

Leaving aside the tautological aspects of their statement, we note here that the presence of a heliconical structure can appear under a variety of local symmetries, over vastly different length scales, and the respective models may be formulated under mutually exclusive conditions and assumptions. Therefore, collecting all models under a single umbrella ($N_{TB}$) does not promote any insights; after all the $N_X$ has specific local symmetry, specific macroscopic symmetry and modulations over a specific length scale. We emphasize in Section II of our reply the serious inconsistencies that arise from an indiscriminate categorization of any nematic that merely shows a heliconical modulation as $N_{TB}$.

The amalgamation of Dozov's and Luckhurst's perspectives may be summarized in the following two statements: i) *There are many $N_{TB}$ models.* ii) *We know for sure that the $N_X$ phase is a $N_{TB}$, but we don't know which $N_{TB}$ it corresponds to*. We conclude that Dozov and Luckhurst seem to place primary importance on the name "$N_{TB}$" and not on the content of the model or the facts about the $N_X$ phase.

## 4. Flawed critique of $N_{PT}$ results and predictions.

In their lengthy dealing with refs DL[14-18] there are a few instances where some of the actual contents are directly criticized by Dozov and Luckhurst. For example, figure 3 of DL[17] was reproduced, with its caption, and then used as a basis for a critique of two aspects of the results. Here we consider them separately and we show that the criticism is wrong.

The temperature dependence of the birefringence is described as the major reason why the $N_{PT}$ model fails:

> "However, this structure is clearly incompatible with some key properties of the $N_X$ phase, e.g. the temperature dependence of the birefringence, which were in perfect agreement with the heliconical $N_{TB}$ structure of the phase."

> "For example, the birefringence is well-known [42] to be proportional to the "coarse-grained" scalar order parameter of the cyanobiphenyl units, i.e. the order parameter $S(L)$ shown in Fig. 3 of Ref. [17] (see this figure reproduced in our Figure 1). The simulated curve is in qualitative disagreement with the precise birefringence data for CB7CB [42] in the "$N_X$" phase – in contrast to the experimentally observed rapid decrease of D$n$ with temperature, $S(L)$ remains constant throughout the "$N_X$" phase. We note that the existing $N_{TB}$ models [26, 41, 42] have been much more successful in explaining the birefringence results."

First, it is not true that that the birefringence in the $N_X$ is proportional to the scalar order parameter of the cyanobiphenyl units (coarse grained or otherwise). As shown experimentally, by direct measurements on CB7CB in Chen et al., (DL[21]; see figure 4), this proportionality holds reasonably well in the $N_U$ but breaks down very clearly in the $N_X$. Therefore, the qualitative disagreement is not between the order parameter in fig. 3 of Ref. DL[17] and the precise birefringence data, but rather between what Dozov and Luckhurst take to be well-known and what is actually measured in DL[21] . Curiously, they never addressed this well-known deviation reported in DL[21]. Rather they extend the proportionality to the $N_X$ phase, as they appear conceptually constrained by the notion of the heliconical nematic director. Moreover, they seem to be unaware of the discussion and possible explanation of precisely this deviation presented in DL[15], the "toy-model": Section 4.3 reads, "This apparently anomalous temperature dependence of the birefringence … could be due to a change in the values of the effective molecular polarizability components as a result of the substantially different averaging of the intermolecular interactions in the N and $N_X$ phases, thus reflecting, albeit indirectly, the effects of the strongly polar ordering."

Second, it is of some interest to see how "perfect agreement" is obtained in their cited work, DL[42]. Examining the section *Theoretical model and discussion* of DL [42] one identifies a whole litany of assumptions on which the calculation is based. Among those are:

(1) The proportionality of anisotropy of the polarizability to the order parameter of the mesogenic units, in both the N and "$N_{TB}$" phase (i.e the same proportionality constant is used for both phases, see eqs(5) and (8)). As discussed above, this is not a valid assumption.

*(2)* The introduction of the conical angle as a parameter. It is this conical angle for which it was later recognized (see section 4; DL[38]) that "…information on the conical angle is not straightforward to obtain and there are significant differences even between the data reported for one given system [14,15,17]. One reason for these differences could be the underlying ambiguity in the definition of the n director."

(3) The introduction of the angle $\psi$, also referred to in Dozov and Luckhurst, section 4.3, as "the angle between the major axis of the mesogenic unit, **L**, and the major axis of the dimer molecule, **z**. Taking into account that $\psi =\sim 30°$…". Such a definition, however, is meaningless for molecules with many different conformations, as is the case with CB7CB. In such case one has a relatively broad, distribution of $\psi$-angles, provided one has consistently defined "the major axis of the dimer molecule, **z**"; this was not done in Ref. DL[42] or, to our knowledge, anywhere else. Apparently, the authors treat the flexible dimer as an effectively rigid molecule.

In summary "perfect agreement" is obtained through a series of questionable assumptions and the introduction of a number of ill-defined adjustable parameters that would enable any model to be brought into perfect agreement with any experimental data.

Following their flawed analysis of the birefringence implications for the $N_{PT}$ Dozov and Luckhurst critique the simulation results for the N-I transition, in particular the non-vanishing values of the order parameter in the I-phase and the lack of a clear discontinuity across the N-I transition:

"However, we do not need to detain ourselves with the details on that point because Fig. 3 of Ref. [17] exhibits even more significant problems with the $N_U$ and I phases which are simpler to simulate… . However, on further heating the behaviour of the order parameter is extremely strange. Across the N-I transition the $S$(L) curve is continuous and without a singularity in its slope, revealing that there is no phase transition at all in this temperature window. In the so-called "isotropic" (I) phase $S$ slowly decreases with increasing temperature from its transitional value at $T_{N-I}$ of 0.25 and remains quite high, $S > 0.08$, for at least 10 °C above the expected N-I "transition". This striking behaviour reveals a serious issue with the simulation of Ref. [17] because its results contradict all of the models, simulations and experimental results for the N-I phase transition (not only for CB7CB, but for *any known nematogen*)."

"Moreover, serious problems in the properties of the simulated phases, e.g. the predicted high, long-range orientational order in the isotropic phase of CB7CB, contradict basic concepts of the theoretical and experimental liquid-crystal physics and undermine completely the numerical results."

Here is what these statements reveal: Initially we note that the simulations of DL[17] are consistent with, i) the established very weak first order nature of the N-I phase transition; and ii) the unavoidable system-size-effects present in any simulation of mesogenic (or other) materials, despite the use of periodic conditions. Those constraints lead to non-vanishing order parameter profiles above the N-I phase transition. This behavior is in general more intense in systems with extended molecular flexibility (e.g., the dimers simulated in DL[17] with nearly atomistic detail) in comparison with the simulations of highly-idealized rigid models. On that topic, consider, for example, the atomistic simulations by Zannoni and his group [15]. Their simulations of the phase behavior of 8CB [15], which presents a similar number of atomistic sites and comparable conformational space volume as CB7CB, show that the variation of the primary orientational order parameter with temperature is continuous across the N-I phase transition (see Fig 2 and Table IV in [15]). Furthermore the values of the order parameter remain quite substantial in the I-phase; the authors specifically point out, "As can be seen in Fig. 2, at high temperatures the sample possesses a very low value of $\langle P_2 \rangle$, ranging from 0.1 to 0.2." Additionally in the same paper they say, "We arbitrarily choose to consider a phase as definitely "ordered" when it shows a $\langle P_2 \rangle$ greater than 0.3, hence locating $T_{NI}$ between 312 and 313 K." Moreover, the authors of DL might wish to consider figure 5 of the simulation paper by Memmer (DL[13]). Do the order parameter values shown therein, exceeding 0.3 in the I phase, "contradict basic concepts of the theoretical and experimental liquid-crystal physics?"

In sum, the "invalidation" of the simulations of Ref. [17] in the Dozov and Luckhurst Comment is totally unfounded. Such arguments reveal their misunderstanding of molecular simulation:

> "As for Memmer's simulations, we expect that for a curved dimer molecule with an anisotropy corresponding to real $N_{TB}$-forming molecules, the calculated pitch will be close to the experimental value of 10 nm."

Their quote above is at odds with the statement by one of them on the same simulations by Memmer DL[12]: "Recent Monte Carlo simulations [8] for achiral banana-shaped molecules also give evidence for temperature-driven transitions from uniform nematic to conical bend-twist helix with macroscopic pitch." In conclusion we reiterate what Memmer's paper states: the pitch length in those simulations is directly related to the box size: $P^* = L_Z^*$.

*5. Excerpts from 3.1 Principles of Continuum Theory*

The conditions for describing the elastic behavior of nematics in the context of the Frank-Oseen continuum theory are specified as follows[10]:

> "For most situations of interest, the-distances $l$ over which significant variations of $Q_{\alpha\beta}$ occur are much larger than the molecular dimensions $a$ (typically $l \sim 1\ \mu m$, while $a \sim 20$Å)

> "Thus the-deformations may be described by a *continuum theory* disregarding the details of the structure on the molecular scale." (p. 98)

> "In a weakly distorted system $(a/l \ll 1)$, at any point, the local optical properties are still those of a uniaxial crystal; the magnitude of the anisotropy is unchanged, it is only the orientation of the optical axis (**n**) that has been rotated. In terms of an order parameter $Q_{\alpha\beta}$

this means that

$$Q_{\alpha\beta} = Q(T)\{n_\alpha(\mathbf{r})n_\beta(\mathbf{r}) - \frac{1}{3}\delta_{\alpha\beta}\} + \text{terms of higher order in } (a/l) \qquad (3.2)$$

The distorted state may then be described entirely in terms of a vector field (**r**). The 'director' **n** is of unit length but of variable orientation. It is assumed that **n** varies slowly and smoothly with **r** (except possibly on a few singular points or singular lines)."

And then, in 3.1.2 "The distortion free energy" (p100):

"We make the following assumptions about this distorted system.
The variations of **n** are slow on the molecular scale

$$a\nabla\mathbf{n} \ll 1."$$

"Let us then call $F_d$ the free energy (per cm$^3$ of nematic material) due to the distortion of **n**. $F_d$ will vanish if $\nabla\mathbf{n} = 0$, and, with our assumptions, it may be expanded in powers of $\nabla\mathbf{n}$. The following conditions must be imposed on $F_d$.

1. $F_d$ must be even in **n**; as explained in Chapter 1, the states (**n**) and (-**n**) are indistinguishable.

2. .... ."

leading to the well-known Frank-Oseen expression for the elastic free energy density (p102):

"Regrouping .... we may write the distortion energy in the form

$$F_d = \frac{1}{2}K_1(\text{div}\,\mathbf{n})^2 \frac{1}{2}K_2(\mathbf{n}\cdot\text{curl}\,\mathbf{n})^2 + \frac{1}{2}K_3(\mathbf{n}\times\text{curl}\,\mathbf{n})^2 \qquad . \qquad (3.15)$$

Equation (3.15) is the fundamental formula of the continuum theory for nematics."

This equation defines the three possible bulk elastic deformations of the nematic director field $\mathbf{n}(\mathbf{r})$, splay, twist and bend; it also specifies their contribution to the free energy density of the deformation.

Finally,

"It is also useful to estimate the magnitude of the distortion energy, per molecule, for a typical distortion taking place in a distance $l$: this will be roughly $F_d a^3 ... \sim U(a/l)^2$. Thus, in the continuum limit ($a \ll l$) it represents only a very small fraction of the total energy." (p. 103)

## 6. History of the Nx phase and its modeling.

The term *twist bend nematic* ($N_{TB}$) originated with Robert Meyer in 1973. His model was based on the continuum elasticity theory of Frank and Oseen, and as such the dimensional scale of its twist modulation was implicitly macroscopic. The second, lower-temperature, nematic phase discovered in cyanobiphenyl dimers (CBnCB) with an odd number, n, of -$CH_2$- units in the aliphatic linker, clearly unrelated to re-entrant nematic phenomena, was commonly denoted $N_X$ since its microscopic structure was unknown. That phase was first observed experimentally by Hiro Toriumi and coworkers in 1991, when Toriumi explicitly suggested that the $N_X$ could be a biaxial nematic. His manuscript "Alkyl Chain Parity Effect in the Phase Transition Behavior of α,ω-Bis(4,4'-cyanobiphenyl)alkane Dimer Liquid Crystals" was submitted to *Liquid Crystals* in August 1991 but Toriumi, discouraged by the review of his paper which requested additional evidence for the biaxiality of the lower temperature nematic phase, abandoned his paper. Concurrently Emerson's and Luckhurst's Preliminary Communication "On the relative propensities of ether and methylene linkages for liquid crystal formation in calamitics" [4] was in press. That paper suggested the "exciting possibility that a biaxial nematic phase could be formed" by the odd-linked CBnCB dimers but was primarily focused on the large even-odd oscillations in thermodynamic attributes of the CBnCB and missed the lower temperature ($N_X$) phase[4]. Two years later Barnes, Douglas, Heeks and Luckhurst abandoned the biaxial idea and concluded that the $N_X$ phase of the n=7 dimer was a smectic phase [5]. It was more than a decade and a half later before the first unequivocal evidence that the $N_X$ phase is a nematic phase was reported in the form of a systematic study of the n=11 dimer made by Panov et al. [6] (SVP[10])). Soon after, enantiotopic discrimination was discovered in the odd-linked CBnCB dimers with NMR implying that a chiral (twisted) arrangement of mesogens must be present in the $N_X$; similar experiments suggested that the upper temperature nematic phase had the characteristics of a uniaxial nematic ($N_U$).

In 2011, Cestari et al. [7] christened the $N_X$ phase *the twist bend nematic* ($N_{TB}$) and since then several hundred papers related to the $N_{TB}$ have appeared, most involving odd-linked CBnCB dimers. These might have constituted corroborative evidence that the $N_{TB}$ had been found at long last, but in 2013 Chen et al. [16] reported that the twist modulation (pitch) observed in the $N_X$ was microscopic ($p = 8.3$ nm). This value of $p$ is not commensurate with the dimensional scale of Meyer's continuum description. In fact that value of $p$ is very close to the length of the crystallographic $c$-axis (8.7 nm) for the "wavy" stacked structure of the n=9 dimer reported by Hori, Iimuri, Nakao and Toriumi [17]. Nevertheless, some continued to try to accommodate both the small value for $p$ and additional experimental evidence at odds with the identification $N_X = N_{TB}$; they did this by expanding the definition of $N_{TB}$ although their criteria for "twist and bend" veered far from Meyer's. As a result, many researchers presently refer to Nx by the misnomer, $N_{TB}$.

In 2016 Vanakaras and Photinos proposed a model of the $N_X$ phase which they named the *polar twisted nematic phase* ($N_{PT}$); it accounted for the extremely tight pitch and other experimental data reported for CBnCB dimers. (It also readily explains the $N_U$ characteristics of the upper temperature nematic if, as suggested by Heist et al. [18], the latter is a cybotactic $N_{PT}$ with a racemic mixture of domains having opposite helicogenic chirality). Dozov and Luckhurst are now expanding the definition of the $N_{TB}$ to encompass a "large family of twist bend nematics," one that includes the Vanakaras/Photinos $N_{PT}$ model (in the small $p$ limit) at one extreme and at the other extreme ($p \rightarrow \infty$) the $N_U$ phase. Such a view relegates the Vanakaras/Photinos model [9] (DL[15]) to the status of a foster child in the $N_{TB}$ family, a superfluous confirmation of all of the preceding modeling. However, the classification of the Nx as a twist bend nematic, a supramolecular structure that derives from Frank-Oseen continuum theory of elasticity in nematics, ignores the fact that the polar twisted nematic, like Frieser's biaxial nematic, derives from mesogen shape and is unconstrained by continuum elastic considerations. It is a molecular model showing how shape polarity (or electrostatic or both) drives a twisted, supramolecular organization. The $N_{PT}$ model has accounted for the small pitch ($p \sim 10$ nm) and all of the previously observed properties of the $N_X$ phase, as well as properties which have been discovered since its introduction.

**Historical omissions in the Invited Comment**

Dozov and Luckhurst omit some of the key episodes mentioned above, but regarding the $N_{TB}$ modeling (*a priori* or *a posteriori*), they say:

> "However, some basic assumptions in these models, the ferroelectric nematic of Meyer, the polar wave condensation of Lorman and Mettout, and the negative bend-distortion modulus of Dozov (implicitly compatible also with the simulations of Memmer) were largely rejected by the scientific community as unrealistic. A notable exception from this generalized disbelief of the $N_{TB}$-phase predictions was a series of papers by Ferrarini and her colleagues [19, 23, 24], in … These studies played a decisive role in stimulating the experimental search of the $N_{TB}$ phase and its discovery first in the CB7CB dimer and later in dozens..."

No references are given to document the supposed 'rejection' by the scientific community, nor are we aware of any reported "disbelief" of the $N_{TB}$ phase predictions. On the other hand, the "notable exception" is well documented.

In their Comment, the section titled "4.1. History of the $N_{PT}$ model and its evolution to match the "$N_X$" phase", Dozov and Luckhurst conclude:

"We revealed that, although the earlier variants of that model [14, 15] proposed a structure for the "$N_X$" phase that was significantly different from the $N_{TB}$ heliconical structure, later, under the growing experimental evidence for the structure of the "$N_X$" phase, their $N_{PT}$ model [16, 17, 18] progressively acquired all the characteristic features previously ascribed to the $N_{TB}$ models, including the heliconical structure, the broken chiral, polar and uniaxial symmetry of this modulated nematic phase. This evolution clearly makes the last variant of the $N_{PT}$ model a new member of the now large family of $N_{TB}$ models, which provides an additional confirmation that the "$N_X$" phase is identical with the predicted $N_{TB}$ phase…"

What is actually revealed here is, i) a misunderstanding of the contents of our Article and of the references DL[14-17]; ii) the attribution of their own problem to us—a model that is changed from paper to paper to fit new data. The latter can be readily identified in a long succession of publications. For example, in 2014 they say,

"The loss of equivalence of deuterons in the $N_{TB}$ phase was interpreted as a change in the phase symmetry, from $D_{\infty h}$ to $D_{\infty}$ " DL[27]

In 2018, the same authors say in DL[38],

"The $N_{TB}$ phase has local $C_2$ symmetry and a local polar director (**m**) parallel to the $C_2$ axis, which rotates in a helical way around a perpendicular axis [4–7]."

Again, in 2014,

" Fig. 1 … scheme of the director organization in the $N_{TB}$ phase: **n** is the director, $p$ is the helical pitch and $\theta_0$ is the conical angle." DL[27]

But in 2018 this is revised,

> "The **n** director, perpendicular to m, can be generally identified with the major average local alignment axis, but unlike **m** it is not uniquely defined [5–7]." DL[38] and,

> "… Actually, information on the conical angle is not straightforward to obtain and there are significant differences even between the data reported for one given system [14,15,17]. One reason for these differences could be the underlying ambiguity in the definition of the n director."

Notably, the figure for the $N_{TB}$ appearing in the Comment by DL shows very clearly the cone angle and the director **n**, despite the abovementioned "underlying ambiguity". It also shows the bend vector, suggesting that the modulation is both a) of elastic orinigin, and b) of molecular dimensions.

In 2015.

> "…similar to the theoretically predicted[11–15] twist-bend phase, $N_{TB}$. In this phase, the bent-shaped <u>mesogenic molecules have long-range orientational order as in usual nematics</u>, and no long-range…." DL[42]

A couple of years later,

> "At the length scale much shorter than the NTB heliconical pitch, $\Lambda \ll p$, the symmetry of the phase is even lower, C2. …… 3b). <u>This symmetry is much lower than the symmetry of the nematic ($D_{\infty h}$)</u>." DL[37]

Finally, in their Comment Dozov and Luckhurst annex current developments reported by other researchers, generating additional confusion:

> "Further, very recently, Chen et al. [70] observed a strong longitudinal ferroelectric polarization <u>in the splay-nematic phase of the same compound</u>. This striking discovery … "

However Chen et al. in their paper state clearly "Our synthesis of RM734 (SI Appendix, section S1) and observation of its electro-optic behavior using polarized light microscopy <u>provides no evidence for a splay nematic phase but rather leads us to the unambiguous conclusion</u> that upon cooling from the higher-temperature, nonpolar, uniaxial nematic (N) phase, RM734 undergoes a transition to <u>another uniaxial nematic ($N_F$) phase that is ferroelectric</u>."